\documentclass[aps,pre,twocolumn,showpacs,amsfonts,amssymb,floatfix]{revtex4}
\usepackage{graphics}
\usepackage{psfrag}
\usepackage{placeins}
\usepackage{graphicx}
\usepackage{amssymb}
\usepackage{amsmath}
\begin{document}

\title{A minimal model for tag-based cooperation}
\author{Arne Traulsen} 
\email{traulsen@theo-physik.uni-kiel.de} 
\author{Heinz Georg Schuster}
\affiliation{Institut f{\"u}r Theoretische Physik und
Astrophysik, Christian-Albrechts Universit{\"a}t,
Olshausenstra{\ss}e 40, 24098 Kiel, Germany}
\date{ \today}

\begin{abstract}

Recently, Riolo et al.\ [R.\ L.\ Riolo et al., Nature 
{\bf 414}, 441 (2001)] showed by computer 
simulations that cooperation can arise  without 
reciprocity when agents donate only to partners 
who are sufficiently similar to themselves. 
One striking outcome of their simulations 
was the observation  that the number of tolerant agents 
that support a wide range of players
was not constant in time, but 
showed characteristic fluctuations. 
The cause and robustness of these 
tides of tolerance remained to be 
explored.
Here we clarify the situation by 
solving a minimal version of the model of 
Riolo et al. It allows us to identify a 
net surplus of random changes from intolerant 
to tolerant agents as a necessary mechanism 
that produces these oscillations of  tolerance 
which segregate different agents in time. This
provides a new mechanism for maintaining different 
agents, i.e.\ for creating biodiversity.
In our model the transition to the oscillating 
state is caused by a saddle node bifurcation.  
The frequency of the oscillations increases 
linearly with the transition rate from tolerant to 
intolerant agents.

\end{abstract}

\maketitle

\section{Introduction}

The emergence of cooperation in evolving populations 
with exploitative individuals is still a challenging
problem in biological and social sciences. Most theories
that explain cooperation are based on direct reciprocity,
as the famous iterated prisoner's dilemma  \cite{Axelrod}.
Cooperation can also arise from indirect reciprocity when
agents help others only if these are known
as sufficiently altruistic \cite{Nowak}.  
In most of these models a finite population of agents is simulated,
pairs of agents meet randomly as potential donator and
receiver. A donation involves some cost to the donor while 
it provides a larger benefit to the receiver. Agents reproduce 
depending on their payoffs after a certain number of such 
meetings. Obviously selfish individuals that do not donate would
quickly spread in the population if help is not channeled 
towards more cooperative players. If agents do not meet
repeatedly---as in a large population---direct 
reciprocity does not work. Indirect reciprocity can 
solve this problems when donations are given only to those individuals 
that are known as sufficiently helpful. This mechanism 
effectively protects a cooperative population against exploiters 
\cite{Nowak}.

Riolo et al.\ \cite{Riolo} introduced a model in 
which cooperation is not based on reciprocity, but on similarity.
In this model donations are channeled towards
individuals that are sufficiently similar to the donator.  
To distinguish between different groups of individuals every agent
$i$ has a tag $\tau_i \in [0,1]$. School ties,
club memberships, tribal costumes or religious creeds are all tags
that induce cooperation. In addition
agents have a tolerance threshold $T_i\geq 0$, which determines
the tag interval that the agent classifies as its own group. 
An agent $i$ donates to
another agent $j$ if their tags are sufficiently
similar, $|\tau_i-\tau_j| \leq T_i$. The cost of
such a donation for $i$ is $c>0$ while the benefit
for $j$ is $b>c$. For simplicity, $b$ is normalized to 1,
since a multiplication of payoffs with a constant factor  
does not change the game. Initially, the tag and 
the tolerance threshold are uniformly
distributed random numbers. 
In each generation every agent acts as a potential
donor for $P$ other agents chosen at random. Hence
it is on average also chosen $P$ times as a recipient.
After each generation each agent $i$ compares its payoff
with the payoff of another randomly chosen agent $j$ and
adopts $T_j$ and $\tau_j$ if $j$ has a higher payoff. In
addition every agent is subject to mutation. With
probability $0.1$ an agent receives a new $\tau$ drawn 
from a uniform distribution
and also with probability $0.1$ a new $T$
which is Gaussian distributed with standard deviation
$\sigma=0.01$ around the old $T$. If this new $T$ becomes 
smaller than zero it is
set to $0$. Obviously, it seems to be the best strategy
for an individual to donate as little as possible, i.e.\ to have a 
very small $T$. However, the whole population would be better off if
everybody would cooperate. This ``tragedy of the commons'' can be solved in different
ways, e.g.\ by volunteering \cite{Hauert1, Hauert2, Szabo}.

\begin{figure}[here]
\begin{center}
\scriptsize
\psfrag{0}{0}
\psfrag{100}{100}
\psfrag{200}{200}
\psfrag{300}{300}
\psfrag{400}{400}
\psfrag{500}{500}
\psfrag{0.00}{0.00}
\psfrag{0.01}{0.01}
\psfrag{0.02}{0.02}
\psfrag{0.03}{0.03}
\psfrag{0.04}{0.04}
\psfrag{0.05}{0.05}
\psfrag{0.20}{0.20}
\psfrag{0.40}{0.40}
\psfrag{0.60}{0.60}
\psfrag{0.80}{0.80}
\psfrag{1.00}{1.00}
\psfrag{Generations}{Generations}
\psfrag{Tolerance}[][][1][180]{Tolerance}
\psfrag{Donation}[][][1][180]{Donationrate}
{\includegraphics[totalheight=3cm]{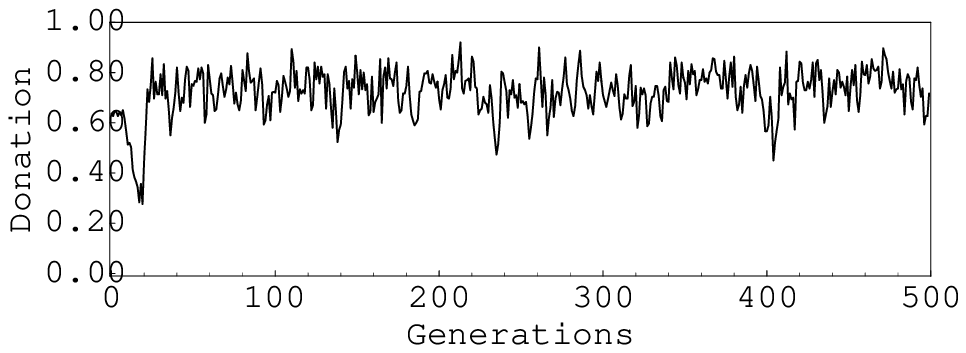}}
{\includegraphics[totalheight=3cm]{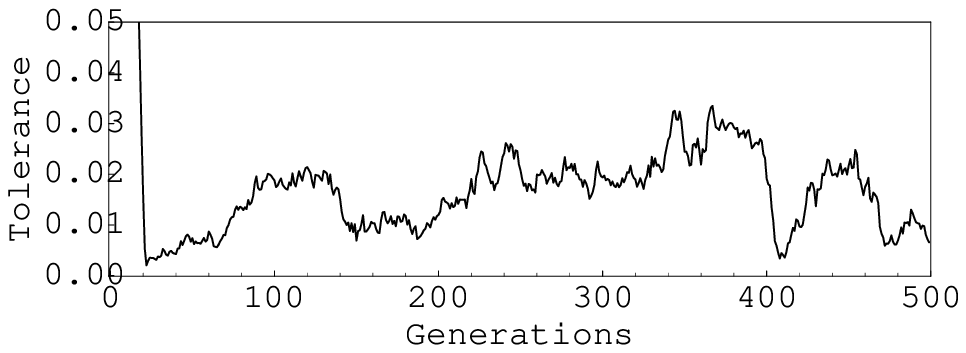}}
\caption{Population dynamics for the first 500 generation
of the model of Riolo et al.\ \cite{Riolo}. 
The average tolerance and the donationrate---i.e.\ the 
fraction of encounters that lead to a donation---show fluctuations. 
When a cooperative cluster becomes dominant its tolerance
increases until the cluster becomes extinct.
($c=0.1$, $b=1.0$, $P=3$).}
\label{Riolo}
\end{center}
\end{figure}

Riolo et al.\ solve this problem by channeling
help towards others that are sufficiently similar to the donator. 
Instead of a cooperative population
the formation and decay of cooperative clusters is
observed for certain parameter ranges (high $P$ and low $c$, see
Fig.\ \ref{Riolo}).
The average tolerance of a cooperative cluster grows slowly over 
time. Occasionally it declines sharply. This decline
occurs when the cluster is exploited by agents that are sufficiently 
similar to the cluster's agents to get support, but do not help themselves.
However, the mechanism that generates these tides of tolerance 
remained unclear \cite{Sigmund}.

Here we develop a minimal model for tag-based cooperation
that displays these ``tides of tolerance'' if
there is a net average drift towards more cooperation.
We find that these fluctuations vanish if such a drift is
not included in the model.
The importance of this observation stems from the 
fact that if we have species that can distinguish 
between themselves and others and donate only to 
others with the same tag, then this would in the 
long run lead to a single group of cooperating 
species having a single tag. But if we introduce a
small rate of biased conversions from intolerant to 
tolerant species we observe a waxing and waning in 
time of species with different tags. In other words, 
the small conversion rate leads to a coexistence of 
different species where different species appear 
cyclically at different times. This consitutes a 
new mechanism that generates biodiversity in a group 
of competing species.

This paper is organized as follows. First the model 
of Riolo et al.\ is simplified in order to allow an analytical 
treatment. Then the system without the effects of mutations 
is analyzed. Thereafter we introduce a drift that increases 
the tolerance and leads to oscillations of tolerance. We 
show that the truncated mutations in the model
of Riolo et al.\ also lead to such a drift.

\section{Simplified replicator model}

\subsection{Definition of the model}

Here we simplify the model of 
Riolo et al. \cite{Riolo} in order to allow for 
an analytical treatment. In a first step we restrict 
the game to only two tags, red and blue. Similarly 
we allow only two tolerances. The agents can either
only donate to others bearing the same tag if they 
have zero tolerance $T=0$ or to every other agent 
($T=1$). This leads to four possible strategies.
Then we allow partners to donate and to receive in
an single interaction instead of defining different 
roles for donators and receivers.
We end up with the payoff matrix
\begin{center}
\begin{tabular}{|l|c|c|c|c|}
\hline  (tag, $T$) & (red, $1$) & (blue, $1$) & (red, $0$) & (blue, $0$) \\
\hline   (red, $1$)  & b-c         & b-c         & b-c         & -c          \\
\hline   (blue, $1$)  & b-c         & b-c         & -c          & b-c         \\
\hline   (red, $0$)  & b-c         & b           & b-c         & 0           \\
\hline   (blue, $0$)  & b           & b-c         & 0           & b-c         \\
\hline
\end{tabular}.
\end{center}
The strategies with $T=1$ are obviously
dominated by the strategies with $T=0$, because the
payoff of an intolerant player is always larger
than the payoff of the corresponding tolerant player. 
There are 
pure Nash equilibria for the intolerant strategies (red, $0$) 
and (blue, $0$). In addition there is an evolutionary 
instable mixed Nash equilibrium if these two strategies 
are used with probability $\frac{1}{2}$. 

If the intolerant agents do not even cooperate within their own group 
we recover the prisoners dilemma \cite{Schusterbook}, 
see Appendix \ref{prisonersdilemma}.

Instead of simulating a finite group of agents we 
calculate only the evolution of the probability 
that an agent uses a certain strategy. 
In the following, $p_1$ and $p_2$ are the frequencies
of tolerant red and tolerant blue agents, respectively. 
$p_3$ and $p_4$ are the frequencies of the corresponding red and blue
intolerant agents. As $p_1^t+p_2^t+p_3^t+p_4^t=1$ the
state of the system is completely determined
by ${\bf p^t}=(p_1^t,p_2^t,p_3^t)$. The trajectory can 
therefore be visualized as a trajectory in the three dimensional
simplex shown in Fig.\ \ref{coordinate}.

In order to apply standard replicator dynamics \cite{Hofbauer} 
we calculate the mean payoffs from the payoff matrix as
\begin{eqnarray}
\label{payoffs}
\Pi_1^t & = & (b - c) (p_1^t + p_2^t + p_3^t) - c\; p_4^t  \nonumber\\
\Pi_2^t & = & (b - c) (p_1^t + p_2^t + p_4^t) - c\; p_3^t \nonumber\\
\Pi_3^t & = & (b - c) (p_1^t + p_3^t) + b\; p_2^t \\
\Pi_4^t & = & (b - c) (p_2^t + p_4^t) + b\; p_1^t \nonumber\\
\langle \Pi  \rangle^t & = & \sum_{i=1}^4 p_i^t \Pi_i^t, \nonumber
\end{eqnarray}
where $\Pi_i$ is the payoff of the strategy with frequency
$p_i$. With (\ref{payoffs}) the replicator equations can be written as
\begin{equation}
{p}_i^{t+1}=p_i^t + p_i^t\; \beta \left(\Pi_i^t -\langle \Pi \rangle^t \right),
\label{RD}
\end{equation}
where $i=1,..,4$. Here $\beta$ determines the time scale.
In the following we set $\beta=1$. Our main interest 
is the attractors of the system, and a
modification of $\beta$ would only modify the velocities
on the attractor.

\subsection{Fixed points and separatrix}

\begin{figure}[here]
\begin{center}
\psfrag{p1}{${\bf p}_1$}
\psfrag{p2}{${\bf p}_2$}
\psfrag{p3}{${\bf p}_3$}
\psfrag{p4}{${\bf p}_4$}
\psfrag{pn}{${\bf p}^n$}
\psfrag{Tp}{${\bf p}^{T+}$}
\psfrag{taup}{${\bf p}^{red}$}
\psfrag{taum}{${\bf p}^{blue}$}
\scriptsize
{\includegraphics[totalheight=7cm]{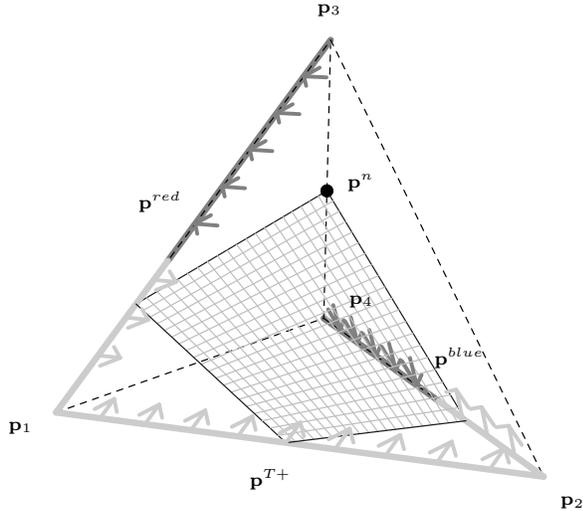}}
\caption{The trajectories of the replicator dynamics move
from the inside of the simplex onto the boundaries. 
The corners represent the pure strategies ${\bf p}_i$.
Arrows indicate the stability of the fixed points 
at the edges. There are two stable
attractors called ${\bf p}^{red}$ and ${\bf p}^{blue}$ (dark grey) 
corresponding to stable lines of fixed points of (\ref{RD}). At the top
only players with red tags survive whereas at the bottom where only players with
blue tags can exist. The two basins of attractions of these 
stable attractors are separated by a planar separatrix
given by (\ref{sepeq}). This separatrix is the basin of attraction
for the fixed point in the Nash equilibrium indicated by a black circle
($c=0.4$, $b=1.0$).}
\label{coordinate}
\end{center}
\end{figure}

The dynamics of the system (\ref{RD}) can roughly be 
characterized as follows, see Fig.\ \ref{coordinate}.
Most initial conditions lead to fixed points where only one tag survives.
The frequency of intolerant players is typically higher 
than the frequency of tolerant players here. 
There is a separatrix that divides the basins of 
attraction of the two tags. On one side
of the separatrix red players will survive and on the 
other side blue players. In addition we find several 
fixed points on the edges described in the following.

As in any replicator system the mixed Nash equilibrium 
${\bf p}^n=(0,0, \frac{1}{2})$ is 
a fixed point. Here the basin of attraction is the 
separatrix. The separatrix shown in 
Fig.\ \ref{coordinate} can be 
calculated from the stability of this fixed point, which is 
discussed for a more general case in Appendix \ref{RDFPStab}. 
${\bf p}^n$ is always part of 
the separatrix, its normal corresponds to the eigenvector 
${\bf e}_3=(1-c,1+c,2)$ of the corresponding Jacobi 
matrix $J^n$ with the eigenvalue $\lambda_3=\frac{3 - c}{2}>1$.
We find the equation 
\begin{equation}
\label{sepeq}
p_3^s=\frac{1}{2}\left[1-(1-c) p_1^s-(1+c) p_2^s\right]
\end{equation}
for points on the separatrix.
As we have $\left({\bf p}^{t+1}(p_1^s,p_2^s,p_3^s)-{\bf p}^{t}(p_1^s,p_2^s,p_3^s)\right) \cdot {\bf e}_3 =0$
the system never leaves this plane again. 

In addition there are two fixed lines if only one tag 
is present: ${\bf p}^{red}=(1-x,0,x)$ and 
${\bf p}^{blue}=(0,1-x,0)$ where $0 \leq x  \leq 1$ is
the fraction of intolerant players.
The stability of the fixed points on these lines
depends on $x$. For $1-x>c$ the points are instable and
intolerant players with the opposite tag can invade 
(see Appendix \ref{RDFPStab}). 
Finally, there is an instable fixed line
for a completely tolerant population, ${\bf p}^{\rm T+}=(1-y,y,0)$, 
where $0\leq y \leq 1$. The stability of this fixed line is 
discussed in Appendix \ref{RDFPStab}.

So far the system does not show any oscillation. It 
simply relaxes to one of the fixed points described
above. In the next section a mechanism that
generates oscillations will be discussed.

\section{Introduction of a biased drift}

In order to generate oscillations in the system we
have to destabilize the attracting fixed points
and force the system through the separatrix. This
can be realized by introducing first ad hoc 
a drift that increases the fraction of tolerant 
agents at the cost of the 
intolerant fraction of the same tag. 
If we introduce such biased conversions 
into our model equation (\ref{RD}) becomes
\begin{eqnarray}
\label{augmentedRD}
{p}_1^{t+1} & = & p_1^t + p_1^t \left(\Pi_1^t -\langle \Pi_t \rangle \right) + \varepsilon p_3^t \\ \nonumber
{p}_2^{t+1} & = & p_2^t + p_2^t \left(\Pi_2^t -\langle \Pi_t \rangle \right) + \varepsilon \left(1-p_1^t-p_2^t-p_3^t \right)\\ \nonumber
{p}_3^{t+1} & = & p_3^t + p_3^t \left(\Pi_3^t -\langle \Pi_t \rangle \right) - \varepsilon p_3^t.
\end{eqnarray}
The solution of these equations shown in Fig.\ \ref{Attractor01}
and  Fig.\ \ref{dynamics} display oscillations in
tolerance. These oscillations can be considered as the deterministic
equivalent to the tides of tolerance in \cite{Riolo}.

\begin{figure}[here]
\begin{center}
\scriptsize
\psfrag{0}{0}
\psfrag{100}{100}
\psfrag{200}{200}
\psfrag{0.00}{0.00}
\psfrag{0.05}{0.05}
\psfrag{0.50}{0.50}
\psfrag{0.10}{0.10}
\psfrag{0.01}{0.01}
\psfrag{1.00}{1.00}
\psfrag{Generations}{Generations}
\psfrag{Tolerance}[][][1][180]{Tolerance}
\psfrag{Donation}[][][1][180]{Donationrate}
{\includegraphics[totalheight=5.5cm]{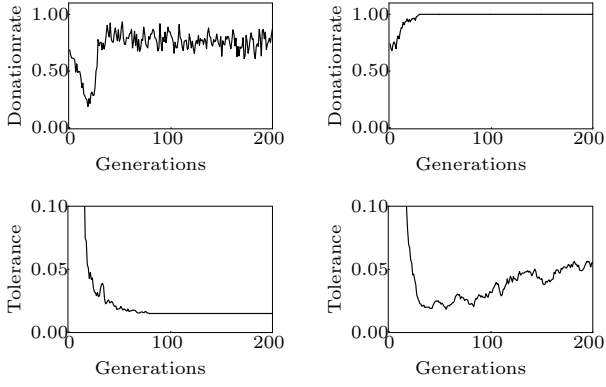}}
\caption{Population dynamics for the first 200 generation
of the model of Riolo et al.\ \cite{Riolo} without tolerance 
mutations (left) and without tag mutations (right).
Without tolerance mutations the donation rate fluctuates due to tag mutations.
All players inherit the same tolerance after less than 100 generations.
Without tag mutations the donation rate quickly raises to $100 \%$ when
all players have the same tag. The
fluctuating tolerance does no longer influence the system
($c=0.1$, $b=1.0$, $P=3$).
}
\label{riolopicnomut}
\end{center}
\end{figure}

In the model of Riolo et al. \cite{Riolo} such a drift 
is generated by truncated mutations.
The average tolerance is usually of the order 
of $\sigma$ Therefore the truncation of negative
tolerances decreases the probability for 
mutations that lower the tolerance and leads to a drift
towards higher tolerances. 
We repeated the simulations
of Riolo et al. and found that $50.0 \%$ of the tolerance
mutations increase T while only $39.8 \%$ decrease T.
The average mutation increases T by $1.3 \cdot 10^{-4}$
($c=0.1$, $P=3$, average over 10 000 realizations with
30 000 generations each). 
If we omit the tolerance mutations in the
model of Riolo et al. one  (low) tolerance
is quickly inherited by the whole populations, see Fig.\ \ref{riolopicnomut}.
The majority of players belongs to a dominant cluster. The
mean tag of this cluster---and hence the donationrate---drifts 
slowly due to mutations of the tags.
Without mutations of the tags one tag is inherited by the whole
population after a short initial period. Consequently the
donation rate becomes $100 \%$, and tolerance mutations
do no longer influence the system. 

\subsection{Qualitative behavior}

\begin{figure}[here]
\begin{center}
\psfrag{p1}{${\bf p}_1$}
\psfrag{p2}{${\bf p}_2$}
\psfrag{p3}{${\bf p}_3$}
\psfrag{p4}{${\bf p}_4$}
\scriptsize
{\includegraphics[totalheight=7cm]{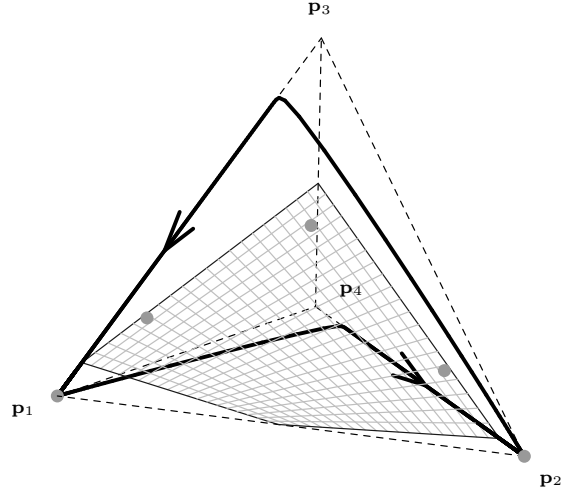}}
\caption{Attractor of the system (\ref{augmentedRD}) for $c=0.1$.
The black line is the attractor, the grey points are
the fixed points. The plane is the separatrix for $\varepsilon=0$.
The arrows indicate how the biased conversions drive the
system through the separatrix to the corner with only tolerant individuals.
Here individuals with the other tag can invade and steer
the system to a corner with mostly intolerant individuals.
Biased conversions lead to a tolerant corner again and the
circle continues ($\varepsilon=0.01$, $c=0.1$, $b=1.0$).}
\label{Attractor01}
\end{center}
\end{figure}

\begin{figure}[here]
\begin{center}
\psfrag{Population density}[][][1][180]{Population density}
\psfrag{Time}{Time}
\psfrag{0.0}{0.0}
\psfrag{0.5}{0.5}
\psfrag{1.0}{1.0}
\psfrag{0}{0}
\psfrag{1000}{1000}
\psfrag{2000}{2000}
\psfrag{3000}{3000}
\psfrag{4000}{4000}
\psfrag{5000}{5000}
\scriptsize
{\includegraphics[totalheight=5cm]{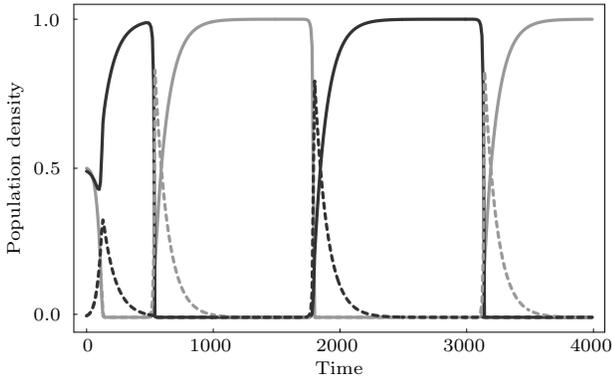}}
\caption{The waxing and waning of the 
four different groups of agents (red agents black, blue agents gray, 
full lines $T=1$, dashed lines $T=0$) 
are caused by the following mechanism. 
A cluster of tolerant red agents 
is invaded by intolerant blue agents  
which convert via directed mutations to 
their tolerant counterpart, giving rise to 
a blue cluster which is then invaded 
by red intolerant agents. Although 
initially the number of red and blue
tolerant agents differed only by one percent 
a tiny number ($0.5\%$) of intolerant agents 
of each tag is enough to generate large 
clusters that are segregated in time 
($\varepsilon=0.01$, $c=0.1$, $b=1.0$).}
\label{dynamics}
\end{center}
\end{figure}

The attractor of the system (\ref{augmentedRD}) is shown in Fig. \ref{Attractor01}, and
the time evolution of the strategies can be seen in Fig.
\ref{dynamics}.
If initially all strategies are present the system shows
periodic oscillations for small $\varepsilon$ and $c=0.1$.
One tag becomes dominant. The fraction of tolerant players
increases due to the biased conversions imposed by $\varepsilon>0$ 
and intolerant players with the opposite tag can invade and destroy the
cluster, giving rise to a new dominant cluster with the opposite tag.
This attractor shown in Fig.\ \ref{Attractor01} has
essentially the whole simplex as a basin of attraction. Only for very small
or very high values of $c$ other fixed points become stable. 
The system can be analyzed in two parts for $\varepsilon  \ll 1$.
Near the edges ${\bf p}^{red}$ and ${\bf p}^{blue}$ the replicator
dynamics becomes irrelevant and the system is mainly driven
by biased conversions. Further away from these edges 
the system is driven by the replicator dynamics. 
Here the dynamics is not altered by the biased conversions.

Our biased conversions lead the system from an edge that 
is dominated by one color to an edge that is dominated by the 
other color. For small $c$ the trajectory leaves these edges
near the corners of the pure tolerant strategies, cf. 
Fig. \ref{Attractor01}. 
However, these corners are never crossed
as they are fixed points.

\subsection{Fixed points}

Let us now analyze the system (\ref{augmentedRD}) in more detail.
The fixed line ${\bf p}^{\rm T+}=(1-y,y,0)$ of (\ref{RD}) is still a
fixed line of (\ref{augmentedRD}). For $c<2\,\varepsilon$ a
fraction of the fixed line remains stable, see Appendix
\ref{RDFPStab} for details. However, as we are 
interested in $\varepsilon \ll 1$ the fixed line is usually instable.
Due to the flow from intolerant to tolerant players
the edges ${\bf p}^{red}$ and ${\bf p}^{blue}$ are no longer fixed.
The fixed point
${\bf p}^n=(0,0, \frac{1}{2})$ in the mixed Nash equilibrium
moves away from the edge for $\varepsilon>0$ and is now given by
${\bf p}^d=(\varepsilon/c,\varepsilon/c, \frac{1}{2}-\varepsilon/c)$.
The stability of this fixed point is discussed in Appendix \ref{RDFPStab}.

In addition we find two more fixed points ${\bf p}^{s+}$ and ${\bf p}^{s-}$.  
For $\varepsilon=0$ they correspond to the
points where the population with only one tag loses stability.
These fixed points can be calculated analytically, see Appendix \ref{AddFPStab}
for details. The expansion for $\varepsilon \ll 1$ of ${\bf p}^{s+}$ is
\begin{equation}
{\bf p}^{s+} \approx \left(\begin{array}{ccc}
1 - c - \frac{2\,\varepsilon }{c} +
  \frac{{\varepsilon }^2}{\left(c -1  \right) \,c^2} \cr
  \frac{{\varepsilon }^2}{c^2 - c^3} \cr
   c + \frac{\left( 1 - 2\,c \right) \,\varepsilon }{c - c^2} -
  \frac{{\varepsilon }^2}{{\left(c -1  \right) }^2\,c}
\end{array}\right).
\end{equation}
Due to the symmetry in the tags ${\bf p}^{s-}$ can
easily be calculated by exchanging $p_1$ with $p_2$
and $p_3$ with $p_4=1-p_1-p_2-p_3$.
As described above we find ${\bf p}^{s+}= (1-c,0,c)$ for $\varepsilon=0$.
Increasing $\varepsilon$ moves it towards ${\bf p}^d$.
For $\varepsilon=c(1-c)/4$ ${\bf p}^{s\pm}$ and ${\bf p}^{d}$ collapse,
here ${\bf p}^{d}$ becomes stable.

For $c<0.73$ we have no fixed points that are stable in all directions.
The whole simplex is essentially the basin of attraction of the attractor shown in 
Fig.\ \ref{Attractor01}.

\subsection{Bifurcation at $\varepsilon=0$}

\begin{figure}[here]
\begin{center}
\scriptsize
\psfrag{p1t}{$p_1^{t}$}
\psfrag{p1t1}[][][1][180]{$p_1^{t+1}$}
\psfrag{0.0}{0.0}
\psfrag{0.2}{0.2}
\psfrag{0.4}{0.4}
\psfrag{0.6}{0.6}
\psfrag{0.8}{0.8}
\psfrag{1.0}{1.0}
\psfrag{A}{{\bf A}}
{\includegraphics[totalheight=7cm]{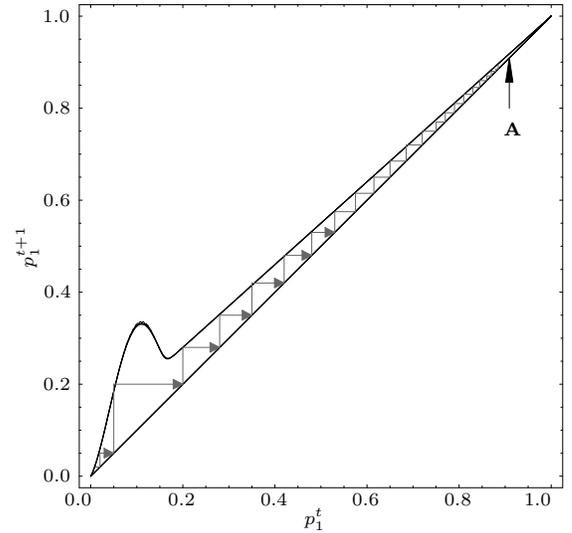}}
\caption{The Poincare map of the $p_1$ 
shows the ``channel'' through which the trajectory 
crosses the separatrix.
The black lines are the function and the bisector. 
The distance between the function and the bisector has been
magnified by a factor of $10$. Therefore the course of iteration is
drawn only schematically. 
{\bf A} marks the point where the 
separatrix is crossed due to biased conversions 
from $p_3$ to $p_1$. Here $p_1$ increases further,
as the fraction $p_4$ that exploits $p_1$ is still very small. 
For $\varepsilon=0$ the function and the bisector will match,
the separatrix can not longer be crossed
($\varepsilon=0.01$, $c=0.1$, $b=1.0$).}
\label{bifurcation}
\end{center}
\end{figure}

The transition from the system without biased conversions, i.e.\
$\varepsilon=0$, to the system with biased conversions can be 
analyzed in detail by considering the Poincare map shown 
in Fig.\ \ref{bifurcation}.

\begin{figure}[here]
\begin{center}
\psfrag{10-1}{$10^{-1}$}
\psfrag{10-2}{$10^{-2}$}
\psfrag{10-3}{$10^{-3}$}
\psfrag{10-4}{$10^{-4}$}
\psfrag{10-5}{$10^{-5}$}
\psfrag{10-6}{$10^{-6}$}
\psfrag{frequency}[][][1][180]{Frequency}
\psfrag{eps}{$\varepsilon$}
\scriptsize
{\includegraphics[totalheight=5cm]{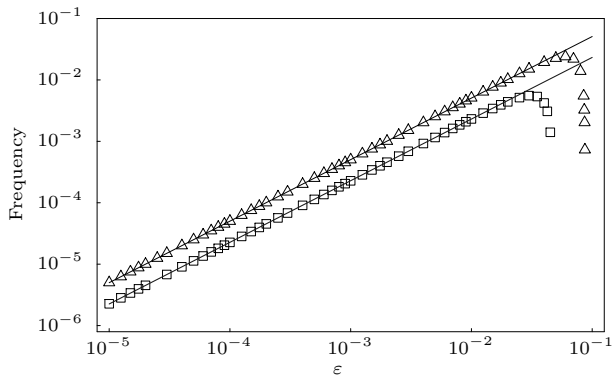}}
\caption{Dependence of the oscillation frequency
on the mutation rate $\varepsilon$. The squares
and the triangles are the numerical values
for $c=0.1$ and $c=0.2$, respectively. 
The line is a fit of
the frequencies for $\varepsilon \leq 0.01$. For small
$\varepsilon$ the frequency increases as
$f=\alpha \varepsilon^{\beta}$. We found
$\beta=1.0036 \pm 0.0003$ for $c=0.1$ and $\beta=1.0021 \pm 0.0002$ for $c=0.2$. 
A linear dependence is expected
if the perturbation is linear in $\varepsilon$,
as in our case. For high values of $\varepsilon$ the fixed line
${\bf p}^{T+}$ becomes partially stable for $\varepsilon=c/2$
and begins to influence the system. Therefore the frequency decreases ($b=1.0$).}
\label{Frequency}
\end{center}
\end{figure}

At $\varepsilon>0$ the fixed lines where only one
tag is present vanish. This is caused by a saddle
node bifurcation \cite{Guckenheimer}.
A fixed line disappears at this bifurcation, and a
small channel is opened through which the system moves slowly
to the other side of the separatrix. The width of this channel 
is controlled by $\varepsilon$. For small 
$\varepsilon$ a linear dependence between $\varepsilon$
and the oscillation frequency of the attractor is observed
as shown in Fig.\ \ref{Frequency}.
Such a linear dependence is expected in a saddle 
node bifurcation with linear perturbation 
terms $\varepsilon p_3$ and $\varepsilon p_4$ \cite{Schusterchaos}.

In our model two small channels are opened by $\varepsilon>0$, 
as the separatrix is crossed twice in one oscillation. 
The reinjection in our model is caused by the
replicator dynamics, which drives the system to
the channel of the opposite tag.
The dependence of the oscillation frequency on
the parameter $\varepsilon$ for $c=0.1$ is shown in
Fig.\ \ref{Frequency}.
For values of $\varepsilon > 0.02$ the dynamic changes. Here the fixed points 
${\bf p}^{T+}$ that become stable for $\varepsilon = c/2$ begin to 
influence the dynamical system.

\subsection{Influence of the cost of cooperation $c$}
\label{cooperationcost}

Here we analyze the
influence of the cost of cooperation $c$ on our system by defining 
different measures of order in our model 
and by observing the influence of $c$ on these measures.
The donation rate is the probability that one player donates to 
another, $d=\langle 1-p_3(p_2+p_4)-p_4(p_1+p_3) \rangle$.
The fraction of tolerant individuals can be measured as 
$p_{\rm tol}=\langle p_1+p_2 \rangle$ ,and the 
asymmetry between the tags as
$a=|\langle p_1 + p_3 \rangle -\langle p_2 + p_4 \rangle|$.
Here $\langle \cdot \rangle$ denotes a time average. In addition an 
average over different initial conditions is necessary.

\begin{figure}[here]
\scriptsize
\begin{center}
\psfrag{0.0}{0.0}
\psfrag{0.2}{0.2}
\psfrag{0.4}{0.4}
\psfrag{0.6}{0.6}
\psfrag{0.8}{0.8}
\psfrag{1.0}{1.0}
\psfrag{c}{cost}
\psfrag{Order parameter}[][][1][180]{Order measure}
{\includegraphics[totalheight=5cm]{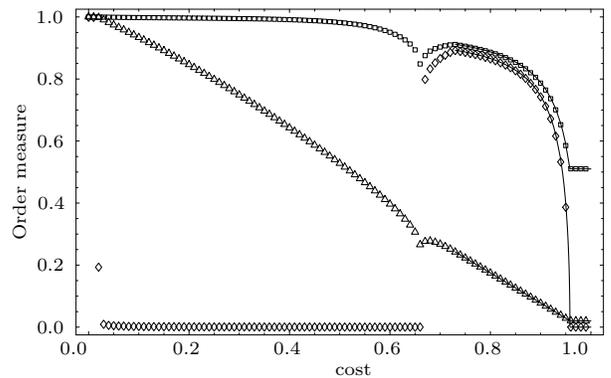}}
\caption{Influence of the cost $c$ on the donation 
rate (squares), the fraction of tolerant players 
(triangles) and the asymmetry between the tags 
(diamonds). All symbols are averages over 10 000 
initial conditions and 100-10 000 time steps. The 
number of time steps is taken as a uniformly distributed random 
number to exclude effects resulting from changes of 
the oscillation frequency. The lines are the analytical 
results for $c>0.73$, see Appendix \ref{AddFPStab}. 
The fraction of tolerant players 
decreases as the time intervals where the tag is invaded 
become longer. This has also an effect on the donation rate. 
For $c \approx 0.66$ a large change of the symmetry 
parameter is observed when one symmetric attractor is 
replaced by two attractors which are not symmetric. 
The fraction of tolerant players and the donation 
rate decrease slightly at $c \approx 0.66$. The donation 
rate and the symmetry parameter increase until the fixed 
points ${\bf p}^{s\pm}$ become stable at $c \approx 0.73$. Here these 
parameters decrease again. When ${\bf p}^{d}$ finally becomes
stable at $c = \frac{1 + {\sqrt{1 - 16\,\epsilon }}}{2} \approx 0.96$
the symmetry is complete again.($\varepsilon=0.01$, $b=1.0$).}
\label{PT}
\end{center}
\end{figure}

Fig.\ \ref{PT} shows that these measures display changes at 
$c \approx 0.02$, $c \approx 0.66$, $c \approx 0.73$, and $c \approx 0.96$.
We now discuss the reasons for these transitions. 
For $c<\varepsilon$ the points ${\bf p}^{T+}=(1-y,y,0)$ 
are stable fixed points. In the case of $\varepsilon<c<2\,\varepsilon$ 
only a part of this fixed line is stable, see Appendix
\ref{RDFPStab} for details. For $c>2\,\varepsilon$ these fixed points 
become instable, this leads to a decrease of the asymmetry between tags
at $c=2\,\varepsilon$ 

\begin{figure}[here]
\begin{center}
\psfrag{p1}{${\bf p}_1$}
\psfrag{p2}{${\bf p}_2$}
\psfrag{p3}{${\bf p}_3$}
\psfrag{p4}{${\bf p}_4$}
\small
{\includegraphics[totalheight=7cm]{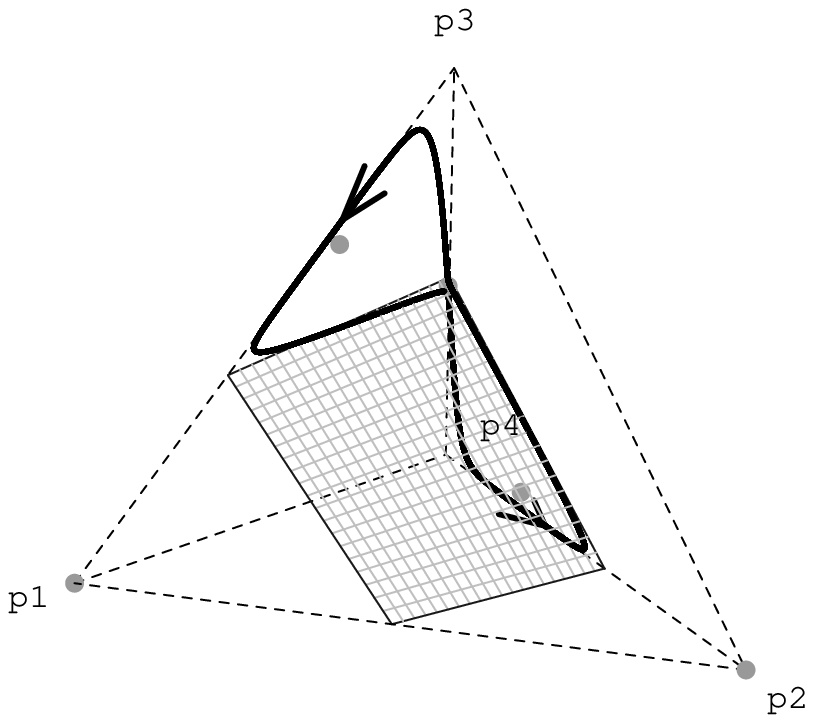}}
\caption{Attractor of the system (\ref{augmentedRD}) for $c=0.66$.
The black line is the attractor. The grey points are
the fixed points. The plane is the separatrix for $\varepsilon=0$.
The arrows indicate the parts of the attractor where it is mainly driven by 
the biased conversions.
The system does no longer cross the separatrix near the edges $p_1+p_3=1$
and $p_2+p_4=1$. Near the fixed point ${\bf p}^d$ the trajectory almost
closes itself. For higher values of $c$ there are two separated attractors
 ( $c=0.66$, $\varepsilon=0.01$, $b=1.0$).}
\label{Attractor066}
\end{center}
\end{figure}

For cooperation costs $c>2\,\varepsilon$ the 
typical qualitative behavior is described above. 
The attractor of such a system can be seen in 
Fig.\ \ref{Attractor01}. For higher costs $c$ 
the intolerant players can invade earlier as their
advantage is larger.
In the following we restrict ourselves to the case 
of $\varepsilon=0.01$. The qualitative behavior 
does not change until $c \approx 0.661$.
The attractor for $c=0.66$ can be seen in Fig.\ \ref{Attractor066}.

For $c > 0.661$ the biased conversion can no longer drive the system
through the separatrix. Two different attractors are observed for
different initial conditions. In the original model this behavior corresponds to the
establishment of one cooperative cluster which becomes tolerant
due to the truncated mutations. Intolerant individuals with the other tag
try to invade, but the dominant cluster becomes more intolerant again and
prevents an invasion.
At $c \approx 0.73$ the fixed points ${\bf p}^{s \pm}$ become stable 
(see Appendix \ref{AddFPStab}). For higher values
of $c$ oscillations are no longer observed. For one
eigenvalue of the corresponding Jacobi matrix $J^s$ we had $|\lambda_1|<1$ even
for smaller $c$. In addition there is a pair of complex conjugated eigenvalues
that crosses the unit circle at $c \approx 0.73$. Hence we are observing a 
Hopf-bifurcation here. For $c>0.73$ the system spins into the fixed points ${\bf p}^{s \pm}$.
For  $c \approx 0.93$ the imaginary parts of the eigenvalues vanish. 
At $c = \frac{1 + {\sqrt{1 - 16\,\epsilon }}}{2} \approx 0.96$ 
the stable fixed points ${\bf p}^{s \pm}$ collapse
with the instable fixed point ${\bf p}^{d}$ in a supercritical pitchfork
bifurcation. For higher values of $c$ the fixed point ${\bf p}^{d}$ is stable.

\section{Summary and Outlook}

We developed a minimal model for cooperation based on similarity. 
This model shows oscillations in 
the population of tolerant agents as two different groups dominate
the population successively. The mechanism that
drives these oscillations is a drift towards more tolerance. 
Without such a drift a cooperative cluster cannot be destabilized
and will not give way to a new cooperative cluster.
In other words, the temporally segregated dynamical coexistence of
different tags is only possible if such a drift towards more 
tolerance exists. Without such a drift only one species would be selected.
This is similar to the dynamical coexistence of species in
the rock-paper-scissors game \cite{Frean}. The drift provides a new mechanism
for maintaining a dynamical biodiversity in biological systems \cite{Kerr}.

This mechanism prevents a single species from taking over the whole
population as it makes the dominant cluster vulnerable.
Agents can therefore exploit the cluster by 
accepting support without supporting the cluster. These free-riders
consequently destroy the cooperative cluster again. 
The cooperative cluster can only defend itself 
if the cost for cooperation is sufficiently high.
In this case the free-riders can not take over the whole population. 

The main results do not change if the 
number of tags is increased. 
However, the analytical treatment becomes much more complicated, as we have to deal
with $n-1$ coupled nonlinear equations in the case of $n$ tags.
Yet, a population model
seems to be more appropriate in the case of more tags, as
our model shows a subsequent realization of all tags in the same order.

If one analyzes a system with a spatial distribution of agents
instead of the well mixed case described above one observes
strong segregation between tags. Tolerant players need to
protect themselves against intolerant exploiters by building a 
border of intolerant agents around them. The spatially
distributed system and the strategies that help to overcome the segregation 
will be discussed in \cite{spatialsegregation}.

\appendix

\section{Prisoners dilemma}
\label{prisonersdilemma}
The introduction of ``never cooperate'' agents which 
do not donate at all \cite{Roberts} instead 
of the zero tolerance agents eliminates the difference 
between tags and leads to the payoff matrix
\begin{center}
\begin{tabular}{|l|c c|c c|}
\hline  (tag, $T$) 	& (red, $+1$) 	& (blue, $+1$) & (red, $0$)& (blue, $0$) 	\\
\hline   (red, $+1$) & b-c         	& b-c         & -c       & -c         	\\
	  (blue, $+1$)& b-c         	& b-c         & -c       & -c       	\\
\hline   (red, $0$)  & b         	& b      	& 0        & 0           	\\
	  (blue, $0$) & b           	& b         	& 0        & 0         	\\
\hline
\end{tabular}
\end{center}
which describes the prisoners dilemma \cite{Axelrod, Schusterbook}.

\section{Fixed points of the Replicator dynamics}
\label{RDFPStab}

The stability of the fixed points with only one tag
can be calculated as follows. For
${\bf p}^{red}=(1-x,0,x)$ and $\varepsilon=0$ we find the Jacobian matrix
\begin{equation} J^{red}=\left(
\begin{array}{ccc} 1+ \left( 1-x \right) \,
     \left( c - x + c\,x \right)    & 0 &
   \left( c-1 \right) \,x^2 \cr
   c - c\,x & 1-x & c\,
   x \cr \left( 1-x \right) \,
     \left( c - x + c\,x \right)    & 0 &
   1+\left( c-1 \right) \,x^2 \cr
   \end{array} \right)
\end {equation}
with the eigenvalues $\lambda_1=1$, $\lambda_2=1-x$ and $\lambda_3=1+c-x$.
The fixed point is marginal stable as long as $x \geq c$, for $x<c$ it becomes
unstable. The reasoning can be adopted for the fixed line ${\bf p}^{blue}=(0,x,0)$.

A fixed point that is conserved for $\varepsilon>0$ 
can be found if all players are tolerant.
For ${\bf p}^{T+}=(1-y,y,0)$ the Jacobian matrix is given by
\begin{equation} J^{T+}=\left(
\begin{array}{ccc}
1+\left(c\,\overline{y}  + y \right) \,\overline{y}  & 
-\overline{c}\, \overline{y} \,y - \varepsilon  & 
0 \cr 
\left( c\,\overline{y}  + y \right) \,\overline{y}  &
1-\overline{c} \,\overline{y} \,y - 
   \varepsilon  & 
0 \cr 
2\, \overline{y}\,y\,\overline{c}+c\,\overline{y} + 
   \varepsilon  & 
-2 \,\overline{y}\,y\,\overline{c}-c\,y - \varepsilon  & 
1+c\,y - 
   \varepsilon  
\end{array} \right)
\end {equation}
where $\overline{y}=1-y$ and $\overline{c}=1-c$.
The eigenvalues of this matrix are
$\lambda_1=1$, $\lambda_2=1+cy-\varepsilon$ and $\lambda_3=1+c(1-y)-\varepsilon$.
$\lambda_i<1$ ($i=1,2,3$) is not possible for $\varepsilon=0$. 
Hence the fixed line is instable for $\varepsilon=0$. 
For $\varepsilon>0$ there
is an interval of stability given by
$1-\frac{\varepsilon}{c}<y<\frac{\varepsilon}{c}$. If this inequation 
and $0 \leq y \leq1$ are both  
fulfilled by $y$, the biased conversions ensure stability of the fixed point
although the replicator dynamics alone would make this point instable.
The first inequation can only be fulfilled for $c<2\, \varepsilon$. For
$c< \varepsilon$ it is always fulfilled and the whole fixed 
line ${\bf p}^{T+}$ is stable.  

The fixed point given by 
${\bf p}^d=(\varepsilon/c,\varepsilon/c,1/2-\varepsilon/c)$
reduces to the mixed Nash equilibrium for $\varepsilon=0$.
The Jacobi matrix at this fixed point is
\begin{equation}
J^d=\left(
\begin{array}{ccc}
	1+\frac{ 3\,c\,\varepsilon +\varepsilon -c^2}{2\,c} & 
    	\frac{- \left( 1 + c \right) \,\varepsilon  }{2\,c} & 
	\frac{ \left( 1 - c \right) \,
       \left( c - 2\,\varepsilon  \right)   }{4\,c} 
\cr 
  	\frac{\left( 1 + c \right) \,\varepsilon }{2\,c} & 
	1+\frac{ c\,\varepsilon - \varepsilon - c^2 }{2\,c} & 
	\frac{\left( 1 + c \right) \,\left( c - 2\,\varepsilon  \right) }{4\,c}
\cr 
	 \frac{\left(1+c\right)\varepsilon}{c}    & 
	\frac{-\left( 1 + c \right) \,\varepsilon }{c}   & 
    	1+\frac{\left( 1 - c \right) \, \left( c - 2\,\varepsilon  \right)   }{2\,c} 
\cr  
\end{array}\right).
\end{equation}
The eigenvalues of this matrix are 
\begin{eqnarray}
\lambda_1 & = & 1-\frac{\gamma}{2}\\
\lambda_2 & = &
	 1+\frac{\gamma (2 c - 1)-
     {\sqrt{ \gamma 
           \left(\gamma + 8 \varepsilon c + 8 \varepsilon c^2 \right) }   }}{4\,c} \nonumber \\	
\lambda_3 & = &
	 1+\frac{\gamma (2 c - 1)+
     {\sqrt{ \gamma 
           \left(\gamma + 8 \varepsilon c + 8 \varepsilon c^2 \right) }   }}{4\,c} \nonumber, 
\end{eqnarray}
where $\gamma=2 \varepsilon -c$.
For $\varepsilon=0$ we have $\lambda_1=\lambda_2=1-\frac{c}{2}<1$ and 
$\lambda_3=\frac{3}{2}-\frac{c}{2}>1$. The third eigenvalue
corresponds to an instable direction. The corresponding eigenvector 
is ${\bf e}_3=(1-c,1+c,2)$, which is the normal of the separatrix
for $\varepsilon=0$. 
In the case of $\varepsilon>0$ we have $\lambda_i<1$ for $i=1,2,3$ only if   
$c>\frac{1 + {\sqrt{1 - 16\,\epsilon }}}{2}$.
Hence ${\bf p}^d$ becomes stable where it coincides with the fixed 
points ${\bf p}^{s+}$ described in appendix \ref{AddFPStab}. In all other cases, 
at least one eigenvalue of $J^d$ is outside the unit circle.

\section{Additional fixed points}

\label{AddFPStab}

Numerical simulations show that the 
additional fixed points for $\varepsilon>0$
can always be found in the plane
spanned by ($1-c,0,c$), ($0,1-c,0$) and ($0,0, \frac{1}{2}$). Together with
$p_1^{t+1}=p_1^t$ and $p_3^{t+1}=p_3^t$ we have three equations that describe these points.
Two of the solutions are fixed points not described above.
The first fixed points ${\bf p}^{s+}$ can be written as
\begin{equation}
{\bf p}^{s+}  = \left(
\begin{array}{ccc}
\frac{ \alpha + {\sqrt{\alpha \beta}} - 2 \varepsilon }{2 c}
\cr
\frac{{\sqrt{\alpha \beta}}  +
    2 \left(  \alpha - \varepsilon  \right)-
     {\sqrt{\alpha^2 + 4 \alpha \beta + 4 \alpha {\sqrt{\alpha \beta}}}}  }{2 c}
\cr
\frac{
    \left( 1-c  \right)\left(c^2 - \beta -2 {\sqrt{\alpha \beta}} \right)
    +{\sqrt{\alpha^2 + 4 \alpha \beta + 4 \alpha {\sqrt{\alpha \beta}}}}}{ 4 \alpha}
\end{array} \right)
\end{equation}
where $\alpha=c(1-c)$ and $\beta=\alpha-4 \varepsilon$.
${\bf p}^{s-}$ can be calculated by exchanging $p_1$ with $p_2$
and $p_3$ with $p_4=1-p_1-p_2-p_3$. These fixed points
have only real coordinates for $\beta \geq 0$.
For $\beta = 0$ we have ${\bf p}^{s+} ={\bf p}^{s-}= {\bf p}^{d}$.

The eigenvalues of the Jacobi matrix at the 
fixed points ${\bf p}^{s\pm}$ can be calculated 
numerically. For $\varepsilon=0.01$ the fixed points are only stable if 
$c>0.73$. At $c = \frac{1 + {\sqrt{1 - 16\,\epsilon }}}{2} \approx 0.96$
they collapse with ${\bf p}^{d}$ in a supercritical pitchfork 
bifurcation and form a single stable fixed point
. 

For $c>0.73$ the fixed points ${\bf p}^{s\pm}$ 
are the only stable attractors and the order measures
described in Section \ref{cooperationcost} 
can be calculated analytically. 
We find for $c<0.96$
\begin{eqnarray}
d & = &  1-p_3(p_2+p_4)-p_4(p_1+p_3) \\ \nonumber
& = & \frac{5 \alpha -4\varepsilon (1+c)+2 \sqrt{\alpha \beta}
-\sqrt{\alpha^2 +4 \alpha \beta +4 \alpha {\sqrt{\alpha \beta}}}}{ 4 \alpha} 
\end{eqnarray}
\begin{eqnarray}
p_{\rm tol} & = & p_1+p_2 \\ \nonumber
& = & \frac{3 \alpha -4\varepsilon+2 \sqrt{\alpha \beta}
-\sqrt{\alpha^2 +4 \alpha \beta +4 \alpha {\sqrt{\alpha \beta}}}}{ 2 c} 
\\
a & = & | p_1+p_3-p_2-p_4 | \\ \nonumber
& = & \frac{-\alpha+ \sqrt{\alpha^2 +4 \alpha \beta +4 \alpha {\sqrt{\alpha \beta}}}}{ 2 \alpha}. 
\end{eqnarray}
For $c> 0.96$ the 
fixed point ${\bf p}^d$ becomes stable and we find
$d  =  \frac{1}{2}+\frac{\varepsilon}{c} $, $p_{\rm tol}  =  2\frac{ \varepsilon}{c}$ and
$a=0$.

\acknowledgments 
\noindent
We thank R. Riolo, M. D. Cohen and 
R. Axelrod for very helpful correspondence and comments. 
A.T. acknowledges support by the  Studienstiftung des 
Deutschen Volkes (German National Merit Foundation).

\small
\bibliographystyle{plain}

\end{document}